\documentstyle[epsf,psfig,referee,supertabular]{mn}
\newcommand{\rhk}{R$^{\prime}_{HK}$}

\newcommand{\Msol}{M$_{\odot}$}

\begin{document}

\title[Anglo-Australian Planet Search]
{Extra-solar planets around HD~196050, HD~216437 and HD~160691}

\author[H. R. A. Jones et al.]{
\parbox[t]{\textwidth}{Hugh R. A. Jones$^1$, R. Paul Butler$^2$, 
%Chris G. Tinney$^3$, Geoffrey W. Marcy$^4$,  
Geoffrey W. Marcy$^3$, Chris G. Tinney$^4$,
Alan J. Penny$^5$, Chris McCarthy$^2$,
Brad D. Carter$^6$} \\
\vspace*{6pt} \\
$^1$Astrophysics Research Institute, Liverpool John Moores University,
Egerton Wharf, Birkenhead CH41 1LD, UK\\
$^2$Carnegie Institution of Washington,Department of Terrestrial Magnetism,
5241 Broad Branch Rd NW, Washington, DC 20015-1305, USA \\
$^3$Department of Astronomy, University of California, Berkeley, CA, 
94720, USA\\
$^4$Anglo-Australian Observatory, PO Box 296, Epping. 1710, Australia \\
$^5$Rutherford Appleton Laboratory, Chilton, Didcot, Oxon OX11 0QX, UK\\
$^6$Faculty of Sciences,  University of Southern Queensland, Toowoomba, 
QLD 4350, Australia\\
}

\date{submitted 29th April 2002}

\maketitle

\label{firstpage}

\begin{abstract}
We report precise Doppler measurements of the stars
HD~216437, HD~196050 and HD~160691 obtained with the 
Anglo-Australian Telescope using the UCLES spectrometer 
together with an iodine cell as part of the 
Anglo-Australian Planet Search.  Our measurements reveal 
periodic Keplerian velocity variations that we interpret as 
evidence for planets in orbit around these solar type stars. 
HD~216437 has a period of 1294$\pm$250~d, a semi-amplitude of 
38$\pm$4 m~s$^{-1}$
and of an eccentricity of 0.33$\pm$0.09. The minimum  
(M~sin~$i$) mass of the companion  is 2.1$\pm$0.3~M$_{\rm JUP}$ and the
semi-major axis is 2.4$\pm$0.5~au. HD~196050 has a period of 1288$\pm$230~d, 
a semi-amplitude of 54$\pm$8~m~s$^{-1}$
and an eccentricity of 0.28$\pm$0.15. The minimum
mass of the companion is 3.0$\pm$0.5~M$_{\rm JUP}$ and the
semi-major axis is 2.3$\pm$0.5~au. 
We also report further observations of the metal rich planet bearing
star HD 160691. Our new solution confirms the previously
reported planet and shows a
trend indicating a second, longer-period companion.
These discoveries add to the growing numbers of midly-eccentric, 
long-period extra-solar planets around Sun-like stars.
As seems to be typical of stars with planets, both
stars are metal-rich. 

\end{abstract}


\section{Introduction}
Radial velocity programmes have now found around 80 extra-solar 
planets orbiting stars in the solar neighbourhood.  As the
time baseline and precision of surveys improve new realms of 
possible planets are being explored. Discoveries include 
the first system of
multiple planets orbiting a Sun-like star (Butler et al. 1999);
the first planet seen in transit (Henry et al. 2000,
Charbonneau et al. 2000); the
first two sub-Saturn-mass planets (Marcy, Butler \& Vogt 2000); 
and the Anglo-Australian Planet Searches' (AAPS)
discovery of the first planet in a circular orbit outside
the 0.1~au tidal-circularisation radius (Butler et al. 2001).
The AAPS began operation in 1998, its southern
hemisphere location completing all-sky coverage 
of the brightest stars at precisions reaching 3~m~s$^{-1}$.
The AAPS has already found 
a number of extra-solar planets (Butler et al. 2001, 2002a; Tinney et al. 2001, 
2002a; Jones et al. 2002). In this paper we present further results from 
this programme. 

\section{The Anglo-Australian Planet Search}

The Anglo-Australian Planet Search (AAPS) is carried out on the 3.9m
Anglo-Australian Telescope using the University College London Echelle
Spectrograph (UCLES), operated in its 31 lines/mm mode 
together with an I$_{2}$ absorption cell. 
UCLES now uses the AAO's EEV 2048$\times$4096 13.5$\mu$m pixel CCD, 
which provides excellent quantum efficiency across the 500--620~nm
I$_2$ absorption line region. 

Doppler shifts are measured by observing through an I$_2$
cell mounted behind the UCLES slit. The resulting superimposed iodine
lines provide a fiducial wavelength scale against which to measure
radial velocity shifts. The shapes of the iodine lines convey the 
PSF of the spectrograph, revealing changes in optics and illumination on 
all time scales. 
Following the procedure of Butler et al. (1996) we synthesize the 
echelle spectrum of each observation on a sub-pixel grid
using a high-resolution reference template, and fit for
spectrograph characteristics (the wavelength scale, scattered light
and the spectrograph PSF) and Doppler shift.
This analysis obtains velocities from multiple epoch observations
measured against a reference template.
This reference template is an observation at the highest
available resolution (using a small 0.5 arcsec slit) and high
signal-to-noise, without the I$_2$ cell present. Such measurements can only 
be efficiently obtained
in good seeing and take about 4 times as long to acquire as a standard
epoch (I$_2$ and a 1 arcsec slit) observation.
Despite this search taking place on a common-user telescope
with frequent changes of instrument, we achieve a 3~m~s$^{-1}$ precision
down to the V~=~7.5 magnitude limit of the survey (Butler et al. 2001; 
fig. 1, Jones et al. 2002).
The fundamental limit to the precision that can be achieved for our sample
is set by a combination of S/N (which is dependent on seeing and
weather conditions), and the intrinsic velocity stability of our target
stars, rather than our observing technique (Butler et al. 1996). 
Intrinsic velocity instability
in these stars -- often called ``jitter'' -- is induced by surface
inhomogeneities due to activity (e.g. spots, plages or flares) 
combined with rotation (Saar et al. 1998; Saar \& Fischer 2000).
There is currently no way to tell whether
a residual scatter of larger than 3~m~s$^{-1}$ is
due to a small-amplitude planet, or jitter induced by
star spots and/or activity.
Only observations over a long enough period to allow the
search for long-term periodicities can reveal the presence of
such relatively small-amplitude long-period signals such as Jupiter.
We intend to monitor all our targets
for the lifetime of the survey, not just those that initially appear to
be good planet candidates.\\

Our target sample which we have observed since 1998 is given in
Table 1. It includes 178 late (IV-V) F, G and K
stars with declinations below $\sim -20^\circ$ and is complete 
to V$<$7.5. We also observe sub-samples of 16 metal-rich 
([Fe/H]$>$0.3) stars with V$<$9.5 and 7 M dwarfs with V$<$7.5 and 
declinations below $\sim -20^\circ$. The sample has been increased 
to around 300 solar-type stars to be complete to a magnitude limit of V=8.
Where age/activity information is
available from R$^{\prime}_{\mathrm HK}$ indices (Henry et al. 1996; Tinney et al. 2002b)
we require target stars to have R$^{\prime}_{\mathrm HK}$ $<$ --4.5 
corresponding to
ages greater than 3 Gyr. Stars with known stellar companions within
2 arcsec are removed from the observing list, as it is operationally difficult
to get an uncontaminated spectrum of a star with a nearby companion.
Spectroscopic binaries discovered during the programme have also been 
removed and are reported by Blundell et al. (2002).
Otherwise there is no bias against observing multiple stars. The programme
is also not expected to have any bias against brown dwarf companions. 
The observing and data processing procedures
follow those described by Butler et al. (1996, 2001).
The first observing run for the AAPS was in 1998 January, and 
the last run for which observations are reported here was in 2002 March.
 
\section{Stellar Characteristics and Orbital Solution for HD~216437}
HD~216437 ($\rho$ Ind, HR\,8701, HIP\,113137) is a
chromospherically inactive
({\rhk}=$-5.01$, Tinney et al. 2002b) G4IV-V star (Cayrel et al. 1997).
Its Hipparcos parallax of 37.7$\pm$0.6\,mas together with a V magnitude
of 6.04 implies an absolute magnitudes
of M$_V$=3.92$\pm$0.03 (ESA 1997) and M$_{\rm bol}$=3.88$\pm$0.03 (Cayrel
et al. 1997).
There is no evidence for significant photometric variability in the 
160 measurements made by the HIPPARCOS satellite.
HD~216437 is known to be somewhat metal-enriched relative to the Sun 
(e.g. [Fe/H]=~0.1, Cayrel de Strobel et al. 1997).
Recent high resolution observations by  Randich et al. (1999) 
have found HD~216437 to have a metallicity of [Fe/H]=~0.21 and 
a lithium abundance of 26~m\AA~ that is consistent
with other similar metal-rich sub-giants. Interpolation
between the tracks of Fuhrmann, Pfeiffer \& Bernkopf (1997,1998) 
indicates a mass
of 1.15$\pm$0.1 for metallicities between solar and [Fe/H]=~0.3.

The 26 Doppler velocity measurements of HD~216437, obtained
between 1998 November and 2002 May, are listed in Table 2 and  
shown graphically in Fig. 1, along with the best fit Keplerian. 
The third column labelled uncertainty
is the velocity uncertainty produced by our least-squares
fitting. 
This uncertainty includes the effects of photon-counting uncertainties,
residual errors in the spectrograph PSF model, and variation in
the underlying spectrum between the template and iodine epochs. All
velocities are measured relative to the zero-point defined by
the template observation. Only observations where the uncertainty
is less than twice the median uncertainty are listed.
The best-fit Keplerian curve
yields an orbital period of $1294\pm250$~d, a velocity amplitude 
of $38\pm4$~m~s$^{-1}$, and
an eccentricity of $0.33\pm0.09$. The minimum (M~sin~$i$) mass of the
planet is $2.1\pm0.3$~M$_{\rm JUP}$, and the semi-major axis is
$2.4\pm0.5$~au. The RMS to the Keplerian fit is 6.64~m~s$^{-1}$, 
yielding a reduced chi-squared of 1.5. The properties of 
the extra-solar planet in orbit around HD~216437 are summarised
in Table 2. 

\section{Stellar Characteristics and Orbital Solution for HD~196050}

HD\,196050 (HIP\,101806) is a chromospherically inactive
({\rhk}=$-5.04$, Henry et al. 1996) G3V star (Houck \& Cowley 1975).
Its Hipparcos parallax of 21.3$\pm$0.9\,mas (ESA 1997) implies absolute 
magnitudes of M$_V$=4.14$\pm$0.05 
and M$_{\rm bol}$=3.94$\pm$0.05 (Drilling \& Landolt 2000).
%
% V=7.501+-, B-V=0.667+-0.01, - pi=21.3+-.9 from HIPPARCOS on-line.
%v-k=1.5, alsono95 roughly mbol =-1 do better?
%Based on its G3V spectral type, one would estimate Teff=5713
%and BC=-0.20 (drilling & landolt), so Mbol=4.25$\pm$0.05
%
% BC(V)=-0.107 Mbol=5.55+-0.05 Teff=6025, so fpb97,fpb98 suggest
%
The fundamental parameters of HD\,196050 have been examined via
B--V and Str\"omgren $ubvy$ photometry (Olsen 1994). These suggest 
T$_{\mathrm eff}$=5590\,K.
Based on interpolation
between the evolutionary tracks by Fuhrmann et al. (1998) 
HD\,196050 is thus estimated to have a metallicity of [Fe/H]=$0.3{\pm}0.2$ 
and a mass of 1.13$\pm$0.1\Msol. 
HD\,196050 is not detected as variable in the
of 144 measurements made by HIPPARCOS. It has recently been used
as an infrared spectroscopic standard by the SOFI instrument on 
the New Technology Telescope at the European Southern 
Observatory in Chile.

The 30 Doppler velocity measurements of HD~196050, obtained
between 1998 November and 2002 May, are listed in Table 3 in 
the same manner as for HD~216437 and
shown graphically in Fig. 2. 
The best-fit Keplerian curve
yields an orbital period of $1288\pm230$~d, a velocity amplitude
of $54\pm8$~m~s$^{-1}$ and
an eccentricity of $0.28\pm0.15$. The minimum (M~sin~$i$) mass of the
planet is 2.3$\pm$0.5~M$_{\rm JUP}$ and the semi-major axis is
3.0$\pm$0.5~au. The RMS to the Keplerian fit is 7.51~m~s$^{-1}$,
yielding a reduced chi-squared of 1.2. The properties of
the extra-solar planet in orbit around HD~196050 are summarised
in Table 2.

\section{A new orbital solution for HD~160691}

We previously announced a companion to HD~160691 (Butler et al.
2001) based on data taken from 1998 November to 2000 November.
Table 3 includes our data up until 2002 May.
All the radial velocities presented in Table 3 have been computed using
an improved template observation of HD~160691 and supercede those
given previously. 
The best-fit single Keplerian curve
yields an orbital period of $638\pm10$~d, a velocity amplitude
of $41\pm5$~m~s$^{-1}$ and
an eccentricity of $0.31\pm0.08$. The minimum (M~sin~$i$) mass of the
planet is 1.7$\pm$0.2~M$_{\rm JUP}$ and the semi-major axis is
1.5$\pm$0.1~au. The RMS to the Keplerian fit is 5.42~m~s$^{-1}$,
yielding a reduced chi-squared of 1.5. The properties of
the extra-solar planet in orbit around HD~160691 are summarised
in Table 2.

The new velocities confirm the planet presented by Butler et al. (2001),
though in addition, Fig. 3 also shows a trend indicating a second companion.
The period of such an outer object is poorly constrained. 
Examination of the parameter space using
the sum of two Keplerians indicates that the RMS is currently
minimized for the ``trend" being
due to an eccentric (0.8) outer planet with a period of 
1300~d and M~sin~$i$ = 1.0~M$_{\rm JUP}$ and an inner planet
with an eccentricity of 0.37 period of 603~d and mass of
1.6~M$_{\rm JUP}$.
However, the data are currently inadequate to provide a convincing case
for this outer planet.  The RMS of the two-planet fit is 4.9~m/s, 
lower than the 5.4~m/s
from the single planet plus linear trend fit, but not statistically
compelling at this time.
Thus these parameters
for the putative outer planet are
speculative pending further velocity measurements.
Any follow-up observations
should take this into account. 
We are mentioning the possibility of this object
at this very early stage in order
that any high precision imaging of HD~160691 may take this 
trend into account.

\section{Discussion}
Although many extra-solar planets had
been discovered prior to 2000 December it was unclear whether
giant planets in circular, or near-circular, orbits outside
0.1~au would be found {\em at all} outside the 
Solar System (e.g. Boss 2001).
The AAPS identification of $\epsilon$\,Ret (Butler et al. 2001)
clearly showed that such planets exist\footnote{
47~Uma was discovered by Butler \& Marcy (1996) but was only realised to 
have a long-period circular orbit when it was discovered
to have two planets in orbit (Fischer et al. 2002)}. Since
our announcement a further 
seven ``$\epsilon$\,Ret-class'' (fig. 4, Tinney et al. 2002a)
have been announced so clearly 
such planets are not as unusual as once thought.  

The newly discovered companions to HD\,216437 and HD\,196050
announced here have masses at least several times that of
Jupiter and have mildly eccentric orbits with periods roughly
twice that of Mars or one-third of Jupiter. These discoveries
serve to reinforce the trend that an     
increasing fraction of the extra-solar planets discovered
have orbital parameters closer to those in our Solar System
than was typical for earlier announcements of extra-solar
planets. In Table 5, we classify the extra-solar planets reported
up until 2002 June.
Around 15\% of extra-solar planetary systems have orbital parameters
within the range of the planets of our Solar System. 
It should be noted that this is probably a lower bound
as we expect that typical orbital parameters and detection frequencies
will evolve considerably as survey baselines and precisions improve. 
Furthermore eccentricity solutions tend to decrease with 
time (Marcy et al. 2002).

\section{Conclusions}

We report extra-solar planets in orbit around the stars    
HD~216437 and HD~196050, and further observations of HD~160691,
which give the preliminary indication of a second planet.
These detections
serve to further emphasize that planetary systems with orbital 
parameters similar to those of our own Solar System are not as 
rare as suggested
by the early extra-solar planet discoveries (e.g., Boss 2001). 
These discoveries confirm the 
preponderance (1) of relatively low-mass 
M~sin~$i$ planets and (2) planets around metal-rich objects.
The detection
of these relatively long-period planets gives us confidence 
in the stability of our search and gives added impetus for the
continuation of the AAPS to longer periods.
We now must endeavour to continue to improve the 
precision and stability of the AAPS to be sensitive
to the 10+ year periods where analogues of the gas giants in 
our own Solar System may become detectable around other 
stars (e.g. Marcy et al. 2002).
 
\section*{Acknowledgments}
The Anglo-Australian Planet Search team would like to thank 
the support of the Director of the AAO, Dr Brian Boyle, 
and the superb technical support which has been received throughout 
the programme from AAT staff -- in particular F.Freeman, D.James,
G.Kitley, S.Lee, J.Pogson, R.Patterson, D.Stafford and J.Stevenson.
We gratefully 
acknowledge the UK and Australian government support of the 
Anglo-Australian Telescope through their PPARC and DETYA funding
(HRAJ, AJP, CGT); NASA grant NAG5-8299 \& NSF grant 
AST95-20443 (GWM); NSF grant AST-9988087 (RPB); and
Sun Microsystems.  This research has made use of the SIMBAD 
database, operated at CDS, Strasbourg, France

\begin{figure}
\psfig{file=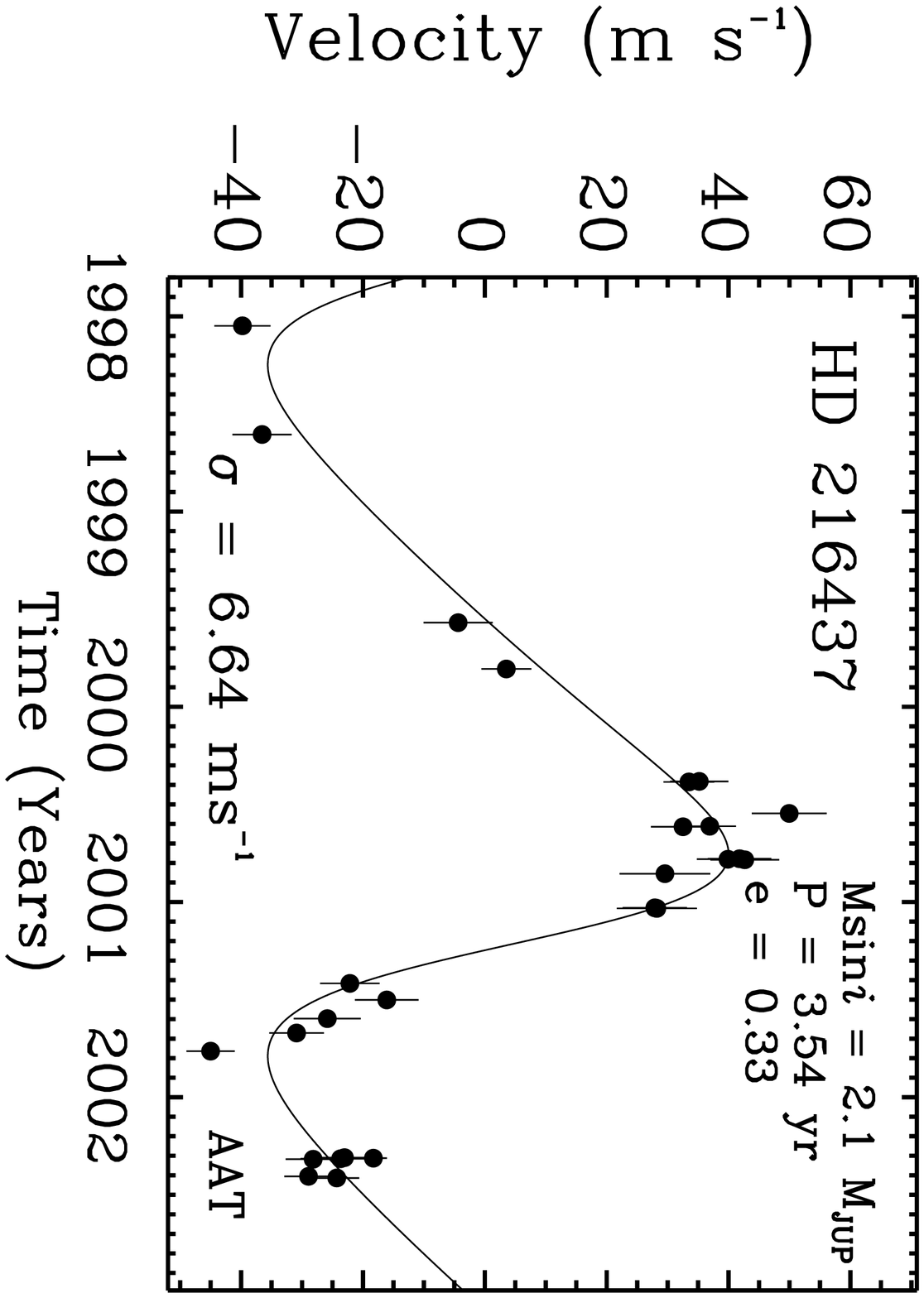,height=8.5in}
\caption{AAT Doppler velocities for HD~216437 from 1998 August to
2001 October. The solid line is a best fit Keplerian orbit with the
parameters shown in Table 2.
The RMS of the velocities about the fit is 6.64~m~s$^{\rm -1}$ 
consistent with our errors.
Assuming 1.15$\pm$0.10~M$_\odot$ for the
primary,
the minimum (M~sin~$i$) mass of the companion is 2.1$\pm$0.3~M$_{\rm JUP}$ and
the semi-major axis is 2.4$\pm$0.5~au.}
\label{hd216437}
\end{figure}

\begin{figure}
\psfig{file=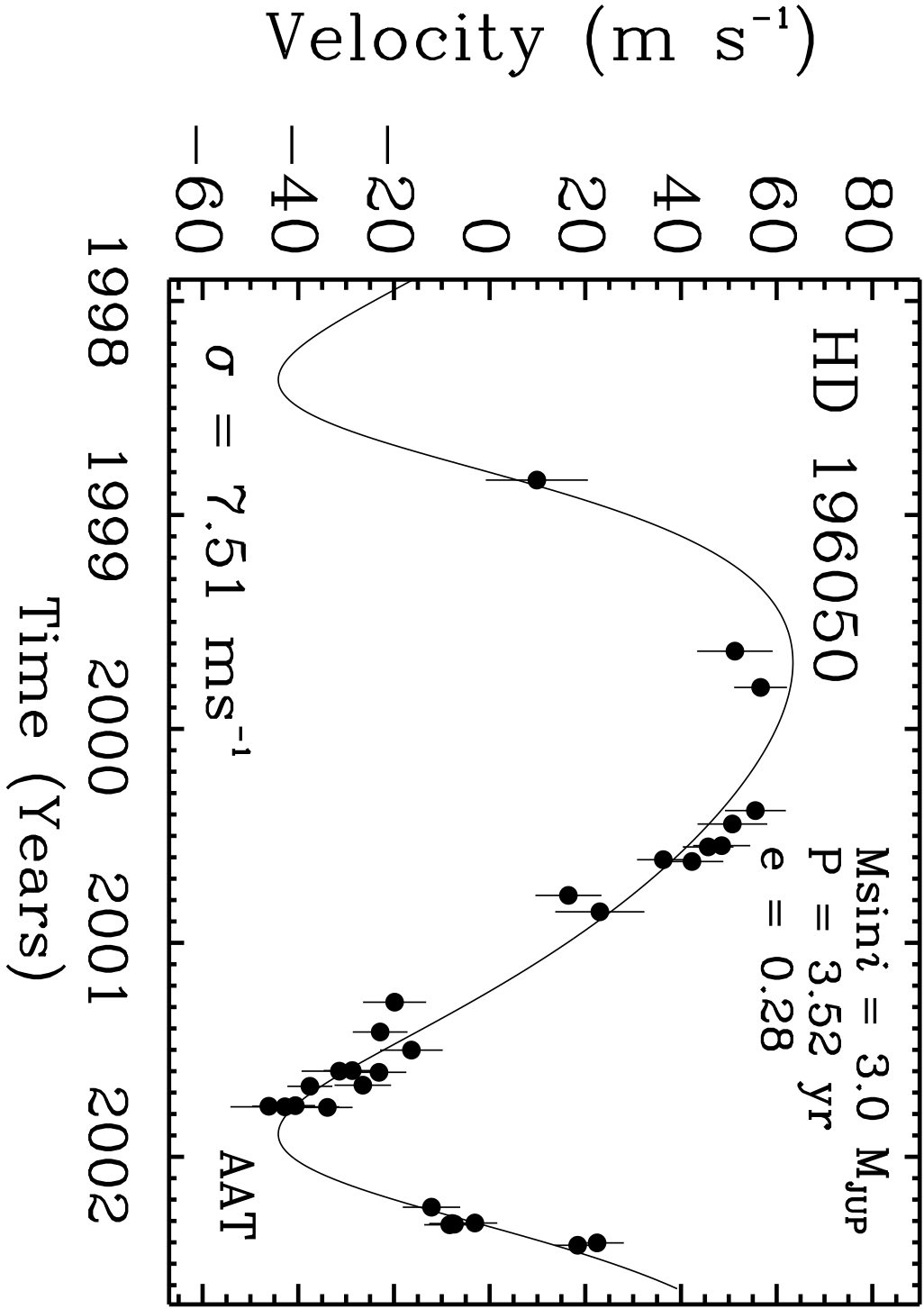,height=8.5in}
\caption{AAT Doppler velocities for HD~196050 from 1998 July to
2002 March. The solid line is a best fit Keplerian orbit with the
parameters shown in Table 2.
The RMS of the velocities about the fit is 7.51~m~s$^{\rm -1}$ 
consistent with our errors.
Assuming 1.13$\pm$0.1 M$_\odot$ for the
primary,
the minimum (M~sin~$i$) mass of the companion is 3.0$\pm$0.5~M$_{\rm JUP}$ and
the semi-major axis is 2.3$\pm$0.5~au.}
\label{hd196050}
\end{figure}

\begin{figure}
\psfig{file=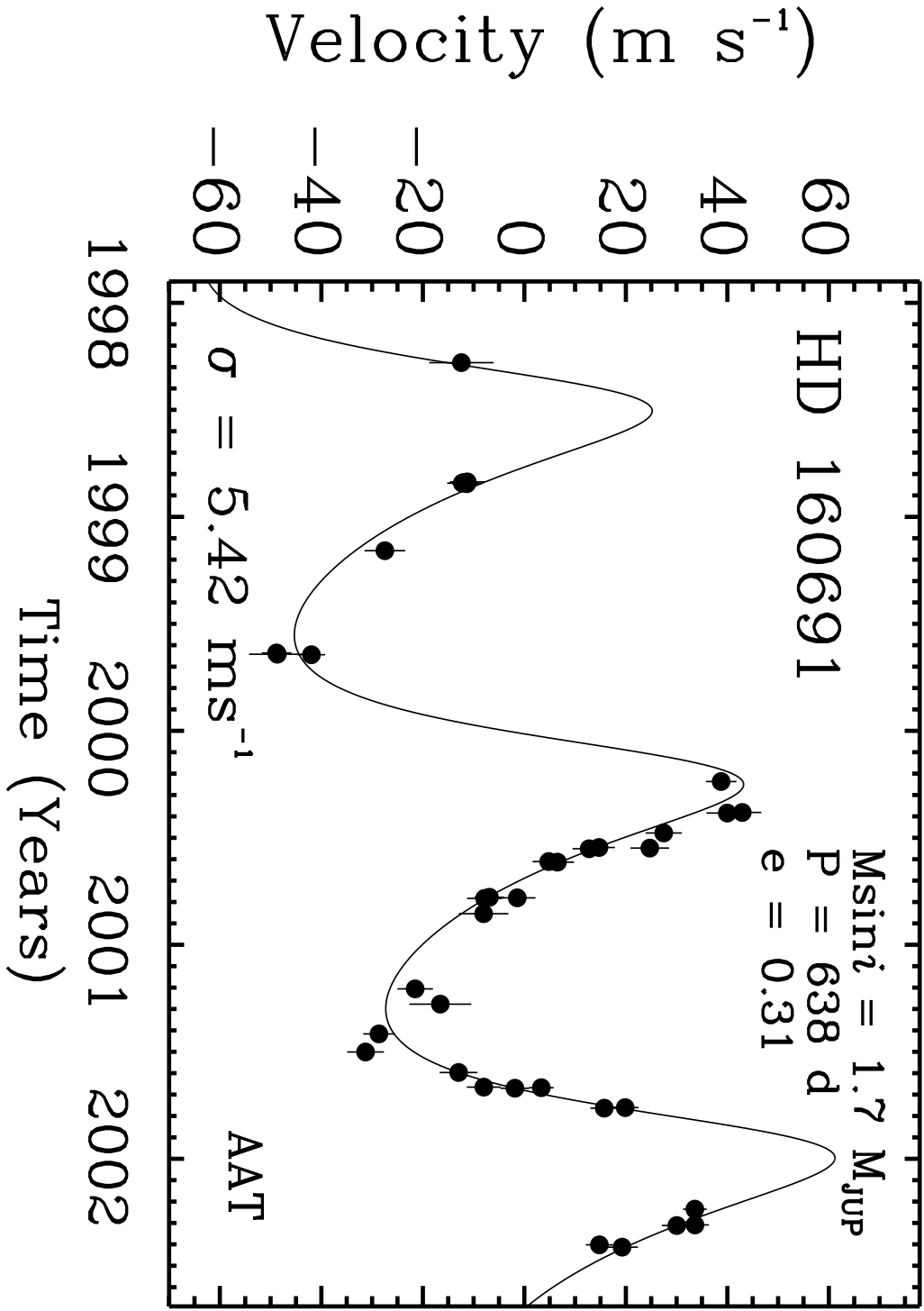,height=8.5in}
\caption{AAT Doppler velocities for HD~160691 from 1998 November to
2002 March. The solid line is a best fit Keplerian orbit with the
parameters shown in Table 2.
The RMS of the velocities about the fit is 5.42~m~s$^{\rm -1}$ 
consistent with our errors.
Assuming 1.08$\pm$0.05~M$_\odot$ for the
primary,
the minimum (M~sin~$i$) mass of the companion is 1.7$\pm$0.2~M$_{\rm JUP}$ and
the semi-major axis is 1.5$\pm$0.1~au.}
\label{hd160691}
\end{figure}

\newpage
 \centering
\begin{table}
\caption{~~Anglo-Australian planet search target list 1998 January to 2002 March.  Doppler companions discovered so far are indicated in the final column.}
\end{table}
\tablefirsthead{HD & RA &Dec & Equinox & V & Sp& Doppler Companion?  \\}
\tablehead{HD&RA &Dec & Equinox & V mag & Sp & Doppler Companion? \\}
\tabletail{}
  \begin{supertabular}{llllllll}

	225213&	00	05	24.2	& -37	21	31	& 2000.0	&	8.56 &		M2V & \\
	142&	00	06	19.0	& -49	04	30	& 2000.0	&	5.70		& G1IV & planet (Tinney et al. 2002a) \\
	1581&	00	20	02.0	& -64	52	39	& 2000.0	&	4.23		& G0V & \\
	2039&	00	24	20.0	& -56	39	00	& 2000.0	&	9.00		& G4V & \\
	2151&	00	25	45.1	& -77	15	15	& 2000.0	&	2.80		& G2IV & \\
	2587&	00	29	10.0	& -50	36	42	& 2000.0	&	8.46		& G7V & \\
	3277&	00	35	34.0	& -39	44	47	& 2000.0	&	7.45		& G6V & star (Blundell et al. 2002) \\
	3823&	00	40	26.4	& -59	27	16	& 2000.0	&	5.89		& G1V & \\
	4308&	00	44	39.0	& -65	38	52	& 2000.0	&	6.55		& G4V & \\
	6735&	01	07	32.0	& -41	44	50	& 2000.0	&	7.01		& F9V & \\
	7199&	01	10	47.0	& -66	11	16	& 2000.0	&	8.06		& K0V & \\
	7570&	01	15	11.0	& -45	31	56	& 2000.0	&	4.97		& G0V & star (Blundell et al. 2002) \\
	9280&	01	31	14.0	& -10	53	48	& 2000.0	&	8.03		& G8V & \\
	10180&	01	37	54.0	& -60	30	41	& 2000.0	&	7.33		& G2V & \\
	10360&	01	39	47.4	& -56	11	53	& 2000.0	&	5.87		& K0V & \\
	10361&	01	39	47.8	& -56	11	41	& 2000.0	&	5.76		& K5V & \\
	10647&	01	42	29.0	& -53	44	26	& 2000.0	&	5.52		& F9V & \\
	10700&	01	44	04.0	& -15	56	15	& 2000.0	&	3.50		& G8V & \\
	11112&	01	48	20.0	& -41	29	43	& 2000.0	&	7.13		& G3V & \\
	12387&	02	00	32.0	& -40	43	51	& 2000.0	&	7.37		& G4V & \\
	13445&	02	10	25.6	& -50	49	28	& 2000.0	&	6.12		& K1V & planet (Butler et al. 2001) \\
	16417&	02	36	58.6	& -34	34	42	& 2000.0	&	5.79		& G5IV & \\
	17051&	02	42	33.2	& -50	48	03	& 2000.0	&	5.40		& G3IV & planet (Butler et al. 2001) \\
	18709&	02	58	59.0	& -43	44	53	& 2000.0	&	7.39		& G1V & \\
	18907&	03	01	37.7	& -28	05	30	& 2000.0	&	5.89		& G5IV & star (Blundell et al. 2002) \\
	19632&	03	08	52.0	& -24	53	17	& 2000.0	&	7.29		& G5V & \\
	20029&	03	11	53.0	& -39	01	23	& 2000.0	&	7.05		& F9V & \\
	20201&	03	12	55.0	& -47	09	20	& 2000.0	&	7.27		& G0V & \\
	20766&	03	17	45.0	& -62	34	37	& 2000.0	&	5.53		& G3V & \\
	20794&	03	19	55.7	& -43	04	11	& 2000.0	&	4.27		& G8V & \\
	20807&	03	18	12.9	& -62	30	23	& 2000.0	&	5.24		& G1V & \\
	20782&	03	20	04.0	& -28	51	13	& 2000.0	&	7.36		& G3V & \\
	22104&	03	27	37.0	& -73	26	24	& 2000.0	&	8.32		& G5V & \\
	23127&	03	39	24.0	& -60	04	42	& 2000.0	&	8.58		& G5V & \\
	23079&	03	39	43.0	& -52	54	57	& 2000.0	&	7.12		& G0V & planet (Tinney et al. 2002a) \\
	23484&	03	44	09.0	& -38	16	54	& 2000.0	&	6.99		& K1V & \\
	24112&	03	48	47.0	& -40	23	58	& 2000.0	&	7.24		& F9V & \\
	25874&	04	02	27.0	& -61	21	26	& 2000.0	&	6.74		& G4V & \\
	25587&	04	02	43.0	& -27	29	00	& 2000.0	&	7.40		& F8V & \\
	26491&	04	07	21.6	& -64	13	21	& 2000.0	&	6.38		& G3V & star (Blundell et al. 2002)\\
	26754&	04	10	07.0	& -61	35	56	& 2000.0	&	7.16		& F9V & \\
	27442&	04	16	28.9	& -59	18	07	& 2000.0	&	4.44		& K2IV & planet (Butler et al. 2001) \\
	28255A&	04	24	12.2	& -57	04	17	& 2000.0	&	6.29		& G4V & \\
	28255B&	04	24	12.2	& -57	04	17	& 2000.0	&	6.60		& G6V & \\
	30177&	04	41	54.0	& -58	01	15	& 2000.0	&	8.41		& G8V & planet (Tinney et al. 2002c) \\
	30295&	04	42	20.0	& -61	37	17	& 2000.0	&	8.86		& G9V & \\
	30876&	04	49	53.0	& -35	06	29	& 2000.0	&	7.49		& K2V & \\
	31527&	04	55	38.0	& -23	14	31	& 2000.0	&	7.49		& G1V & \\
	31827&	04	56	18.0	& -51	02	50	& 2000.0	&	8.26		& G8V & \\
	33811&	05	10	43.0	& -44	34	20	& 2000.0	&	8.71		& G8V & \\
	36108&	05	28	21.0	& -22	26	04	& 2000.0	&	6.78		& G1V & \\
	38283&	05	37	02.0	& -73	41	58	& 2000.0	&	6.69		& G0V & \\
	39091&	05	37	09.8	& -80	28	09	& 2000.0	&	5.65		& G1V & planet (Jones et al. 2002) \\
	38110&	05	42	59.0	& -07	28	51	& 2000.0	&	8.18		& G5V & \\
	38382&	05	44	28.0	& -20	07	35	& 2000.0	&	6.34		& G0V & \\
	38973&	05	46	28.0	& -53	13	09	& 2000.0	&	6.63		& G1V & \\
	39213&	05	49	16.0	& -37	30	48	& 2000.0	&	8.96		& G9V & star (Blundell et al. 2002) \\
	40307&	05	54	04.0	& -60	01	24	& 2000.0	&	7.17		& K2V & \\
	42024&	06	06	12.0	& -45	48	58	& 2000.0	&	7.24		& F9V & star (Blundell et al. 2002) \\
	43834&	06	10	14.4	& -74	45	11	& 2000.0	&	5.09		& G6V & \\
	42902&	06	11	14.0	& -44	13	28	& 2000.0	&	8.92		& G2V & \\
	44447&	06	15	06.0	& -71	42	10	& 2000.0	&	6.62		& F9V & \\
	44120&	06	16	18.5	& -59	12	49	& 2000.0	&	6.43		& G0V & \\
	44594&	06	20	06.0	& -48	44	26	& 2000.0	&	6.61		& G4V & \\
	45289&	06	24	24.0	& -42	50	28	& 2000.0	&	6.67		& G5V & \\
	45701&	06	24	26.0	& -63	25	44	& 2000.0	&	6.45		& G4V & \\
	52447&	06	57	26.0	& -60	51	05	& 2000.0	&	8.38		& G1V & \\
	53705&	07	03	57.3	& -43	36	29	& 2000.0	&	5.54		& G3V & \\
	53706&	07	03	59.0	& -43	36	44	& 2000.0	&	6.83		& G8V & \\
	55720&	07	11	32.0	& -49	25	29	& 2000.0	&	7.50		& G6V & \\
	55693&	07	13	03.0	& -24	13	33	& 2000.0	&	7.17		& G4V & \\
	59468&	07	27	26.0	& -51	24	09	& 2000.0	&	6.72		& G5V & \\
	61686&	07	39	35.0	& -26	28	28	& 2000.0	&	8.54		& G5V & \\
	64184&	07	49	27.0	& -59	22	52	& 2000.0	&	7.49		& G5V & star (Blundell et al. 2002) \\
	65907A&	07	57	46.9	& -60	18	12	& 2000.0	&	5.60		& G0V & \\
	67199&	08	02	31.0	& -66	01	18	& 2000.0	&	7.18		& K1V & \\
	67556&	08	07	09.0	& -36	22	54	& 2000.0	&	7.30		& F8V & \\
	69655&	08	15	26.0	& -52	03	37	& 2000.0	&	6.63		& G0V & \\
	70642&	08	21	28.0	& -39	42	21	& 2000.0	&	7.17		& G5V & \\
	70889&	08	23	32.0	& -27	49	21	& 2000.0	&	7.09		& G1V & \\
	72769&	08	33	46.0	& -23	21	18	& 2000.0	&	7.22		& G7V & \\
	73121&	08	35	12.6	& -39	58	12	& 2000.0	&	6.47		& G1V & \\
	73526&	08	37	17.0	& -41	19	10	& 2000.0	&	8.99		& G7V & planet (Tinney et al. 2002c) \\
	73524&	08	37	20.0	& -40	08	51	& 2000.0	&	6.53		& G1V & \\
	74868&	08	44	51.0	& -44	32	34	& 2000.0	&	6.56		& F9V & \\
	75289&	08	47	41.0	& -41	44	14	& 2000.0	&	6.35		& G0V & planet (Butler et al. 2001) \\
	76700&	08	53	54.0	& -66	48	05	& 2000.0	&	8.16		& G7V & \\
	78429&	09	06	39.0	& -43	29	32	& 2000.0	&	7.31		& G4V & \\
	80913&	09	12	26.0	& -81	46	08	& 2000.0	&	7.49		& F9V & \\
	80635&	09	20	27.0	& -17	25	29	& 2000.0	&	8.80		& G6V & \\
	82082&	09	27	32.0	& -58	05	40	& 2000.0	&	7.20		& G1V & \\
	83443&	09	37	12.0	& -43	16	19	& 2000.0	&	8.23		& G9V & planet (Butler et al. 2002a) \\
	83529A&	09	37	29.0	& -49	59	27	& 2000.0	&	6.97		& G0V & \\
	84117&	09	42	15.0	& -23	54	58	& 2000.0	&	4.93		& F8V & \\
	85683&	09	51	41.0	& -54	39	35	& 2000.0	&	7.34		& F8V & \\
	86819&	10	00	06.0	& -36	02	36	& 2000.0	&	7.38		& G0V & \\
	88742&	10	13	25.0	& -33	01	55	& 2000.0	&	6.38		& G1V & \\
	92987&	10	43	36.0	& -39	03	31	& 2000.0	&	7.03		& G3V & \\
	93385&	10	46	15.0	& -41	27	52	& 2000.0	&	7.49		& G1V & \\
	96423&	11	06	20.0	& -44	22	24	& 2000.0	&	7.23		& G5V & \\
	101614&	11	41	27.0	& -41	01	06	& 2000.0	&	6.86		& G1V & \\
	101959&	11	43	57.0	& -29	44	51	& 2000.0	&	6.97		& F9V & \\
	102117&	11	44	50.0	& -58	42	12	& 2000.0	&	7.47		& G6V & \\
	102365&	11	46	31.1	& -40	30	02	& 2000.0	&	4.91		& G3V & \\
	102438&	11	47	15.7	& -30	17	13	& 2000.0	&	6.48		& G5V & \\
	105328&	12	07	39.0	& -23	58	33	& 2000.0	&	6.72		& G2V & \\
	106453&	12	14	42.0	& -24	46	34	& 2000.0	&	7.47		& G6V & \\
	107692&	12	22	45.0	& -39	10	38	& 2000.0	&	6.70		& G3V & \\
	108147&	12	25	46.0	& -64	01	22	& 2000.0	&	6.99		& F8V & planet (Pepe et al. 2002) \\
	108309&	12	26	48.2	& -48	54	48	& 2000.0	&	6.26		& G3-5V & \\
	109200&	12	33	32.0	& -68	45	20	& 2000.0	&	7.13		& K0V & \\
	114613&	13	12	03.2	& -37	48	11	& 2000.0	&	4.85		& G3V & \\
	114853&	13	13	52.0	& -45	11	10	& 2000.0	&	6.93		& G3V & \\
	117618&	13	32	26.0	& -47	16	18	& 2000.0	&	7.17		& G1V & \\
	118972&	13	41	04.0	& -34	27	50	& 2000.0	&	6.92		& K0V & \\
	120237&	13	48	55.0	& -35	42	14	& 2000.0	&	6.56		& F9V & \\
	120690&	13	51	20.0	& -24	23	27	& 2000.0	&	6.43		& G6V & star (Blundell et al. 2002) \\
	121384&	13	56	33.0	& -54	42	16	& 2000.0	&	6.00		& G6IV-V & star (Blundell et al. 2002) \\
	122862&	14	08	27.1	& -74	51	01	& 2000.0	&	6.02		& G2-3IV & \\
	125072&	14	19	05.0	& -59	22	37	& 2000.0	&	6.66		& K4V & \\
	GL551& 14	29	42.2	& -62	40	48	& 2000.0	&	11.01 &		M5V & \\
	128620	&14	39	35.9	& -60	50	07	& 2000.0	&	-0.01		& G2V & \\
	128621&	14	39	36.1	& -60	50	08	& 2000.0	&	1.33		& K1V & \\
	129060&	14	44	14.0	& -69	40	28	& 2000.0	&	6.99		& F9V & \\
	131923&	14	58	08.8	& -48	51	47	& 2000.0	&	6.35		& G3-5V & star (Blundell et al. 2002) \\
	134331&	15	10	42.0	& -43	43	48	& 2000.0	&	7.01		& G2V & \\
	134330&	15	10	43.0	& -43	42	58	& 2000.0	&	7.60		& G6V & \\
	134060&	15	10	44.6	& -61	25	21	& 2000.0	&	6.30		& G2V & \\
	134987&	15	13	28.7	& -25	18	33	& 2000.0	&	6.45		& G4V & planet (Butler et al. 2001) \\
	134606&	15	15	15.0	& -70	31	11	& 2000.0	&	6.86		& G7V & \\
	136352&	15	21	48.2	& -48	19	04	& 2000.0	&	5.65		& G3-5V & \\
	140901&	15	47	29.0	& -37	54	59	& 2000.0	&	6.01		& G6V & \\
	143114&	15	59	38.0	& -29	37	58	& 2000.0	&	7.34		& G1V & \\
	144628&	16	09	43.0	& -56	26	43	& 2000.0	&	7.11		& K0V & \\
	145825&	16	14	12.0	& -31	39	47	& 2000.0	&	6.55		& G3V & star (Blundell et al. 2002) \\
	147722&	16	24	39.6	& -29	42	12	& 2000.0	&	6.50		& G0IV & \\
	147723&	16	24	39.7	& -29	42	17	& 2000.0	&	5.84		& G0IV & \\
	150248&	16	41	50.0	& -45	22	07	& 2000.0	&	7.03		& G4V & star (Blundell et al. 2002) \\
	154577&	17	10	11.0	& -60	43	42	& 2000.0	&	7.38		& K1V & \\
	155974&	17	16	21.5	& -35	44	58	& 2000.0	&	6.12		& G0IV-V & \\
	156274A&17	19	03.0	& -46	38	13	& 2000.0	&	7.0:	&	M0V & star (Blundell et al. 2002) \\
	156274B	&17	19	04.3	& -46	38	10	& 2000.0	&	5.52		& K0V & \\
	158783&	17	34	12.0	& -54	53	43	& 2000.0	&	7.09		& G4V & star (Blundell et al. 2002) \\
	160691&	17	44	08.7	& -51	50	03	& 2000.0	&	5.15		& G3IV-V & planet (Butler et al. 2001; this paper) \\
	161050&	17	47	46.0	& -63	33	45	& 2000.0	&	7.16		& G1V & \\
	161612&	17	47	57.0	& -34	01	07	& 2000.0	&	7.20		& G7V & \\
	162255&	17	51	08.0	& -22	55	14	& 2000.0	&	7.15		& G3V & star (Blundell et al. 2002) \\
	164427&	18	04	43.0	& -59	12	36	& 2000.0	&	6.88		& G2V & brown dwarf (Tinney et al. 2001) \\
	168871&	18	24	33.0	& -49	39	10	& 2000.0	&	6.45		& G1V & \\
	169586&	18	26	41.0	& -30	23	37	& 2000.0	&	6.75		& F8V & star (Blundell et al. 2002) \\
	GL729&18	49	49.0	& -23	50	10	& 2000.0	&	10.46 & M4V & \\
	175345	&18	56	00.0	& -25	02	48	& 2000.0	&	7.37		& F9V & star (Blundell et al. 2002) \\
	177565	&19	06	52.5	& -37	48	37	& 2000.0	&	6.16		& G5IV & \\
	179949	&19	15	33.0	& -24	10	45	& 2000.0	&	6.25		& F8V & planet (Tinney et al. 2001) \\
	181428	&19	21	39.0	& -29	36	19	& 2000.0	&	7.10		& F9V & \\
	183877	&19	32	40.0	& -28	01	11	& 2000.0	&	7.14		& G5V & \\
	187085	&19	49	34.0	& -37	46	50	& 2000.0	&	7.22		& G0V & \\
	189567	&20	05	32.8	& -67	19	15	& 2000.0	&	6.07		& G3V & \\
	190248	&20	08	43.6	& -66	10	55	& 2000.0	&	3.56		& G6-8IV & \\
	191408	&20	11	11.9	& -36	06	04	& 2000.0	&	5.32		& K3V & \\
	192310	&20	15	17.4	& -27	01	58	& 2000.0	&	5.73		& K0V & \\
	193193	&20	19	45.0	& -25	13	43	& 2000.0	&	7.20		& G1V & \\
	192865	&20	21	36.0	& -67	18	46	& 2000.0	&	6.91		& F9V & \\
	193307	&20	21	41.0	& -49	59	58	& 2000.0	&	6.27		& G0V & \\
	194640	&20	27	44.0	& -30	52	00	& 2000.0	&	6.61		& G6V & \\
	196050	&20	37	52.0	& -60	38	03	& 2000.0	&	7.50		& G4V & planet (this paper)\\
	196800	&20	40	22.0	& -24	07	04	& 2000.0	&	7.21		& G2V & \\
	196068	&20	41	45.0	& -75	20	46	& 2000.0	&	7.18		& G3V & \\
	196378	&20	40	02.3	& -60	32	51	& 2000.0	&	5.11		& F8V & \\
	199288	&20	57	40.0	& -44	07	37	& 2000.0	&	6.52		& G0V & \\
	199190	&21	00	06.0	& -69	34	45	& 2000.0	&	6.86		& G3V & \\
	199509	&21	09	22.0	& -82	01	37	& 2000.0	&	6.98		& G2V & \\
	202560&	21	17	15.0	& -38	52	04	& 2000.0	&	6.69		&M0V & \\
	202628&	21	18	27.0	& -43	20	05	& 2000.0	&	6.75		& G3V & \\
	204385&	21	30	48.0	& -62	10	06	& 2000.0	&	7.14		& G1V & \\
	204961&	21	33	34.0	& -49	00	25	& 2000.0	&	8.66 &		G1V & \\
	205390&	21	36	41.0	& -50	50	46	& 2000.0	&	7.15		& K1V & \\
	205536&	21	40	31.0	& -74	04	28	& 2000.0	&	7.07		& G7V & \\
	206395&	21	43	02.0	& -43	29	46	& 2000.0	&	6.67		& F9V & \\
	207129&	21	48	15.8	& -47	18	13	& 2000.0	&	5.58		& G0V & \\
	207700&	21	54	46.0	& -73	26	17	& 2000.0	&	7.43		& G5V & \\
	208487&	21	57	20.0	& -37	45	52	& 2000.0	&	7.47		& F9V & \\
	208998&	22	01	37.0	& -53	05	36	& 2000.0	&	7.12		& G0V & \\
	209268&	22	03	35.0	& -55	58	38	& 2000.0	&	6.88		& F9V & \\
	209653&	22	07	31.0	& -68	01	23	& 2000.0	&	6.99		& G0V & \\
	210918&	22	14	38.6	& -41	22	54	& 2000.0	&	6.23		& G5V & star (Blundell et al. 2002) \\
	211317&	22	18	50.0	& -68	18	47	& 2000.0	&	7.26		& G4V & \\
	212330&	22	24	56.4	& -57	47	50	& 2000.0	&	5.32		& G3IV & \\
	212168&	22	25	51.0	& -75	00	56	& 2000.0	&	6.04		& G3V & \\
	212708&	22	27	25.0	& -49	21	58	& 2000.0	&	7.48		& G7V & \\
	213240&	22	31	00.0	& -49	26	00	& 2000.0	&	6.81		& G1V & planet (Santos et al. 2001) \\
	214759&	22	40	55.0	& -31	59	23	& 2000.0	&	7.41		& G8V & \\
	214953&	22	42	36.9	& -47	12	38	& 2000.0	&	5.98		& G0V & \\
	216435&	22	53	37.9	& -48	35	53	& 2000.0	&	6.04		& G0V & \\
	216437&	22	54	39.4	& -70	04	25	& 2000.0	&	6.05		& G2-3IV & planet (this paper)\\
	217958&	23	04	33.0	& -25	41	27	& 2000.0	&	8.05		& G4V & \\
	217987&	23	05	51.2	& -35	51	11	& 2000.0	&	7.35		& M2V & \\
	219077&	23	14	06.6	& -62	42	00	& 2000.0	&	6.12		& G8V & \\
	220507&	23	24	42.0	& -52	42	08	& 2000.0	&	7.59		& G5V & \\
	221420&	23	33	19.5	& -77	23	07	& 2000.0	&	5.81		& G2V & \\
	222237&	23	39	37.0	& -72	43	19	& 2000.0	&	7.09		& K3V & \\
	222335&	23	39	51.0	& -32	44	34	& 2000.0	&	7.18		& G9V & \\
	222480&	23	41	08.0	& -32	04	14	& 2000.0	&	7.11		& G4V & \\
	223171&	23	47	21.0	& -48	16	33	& 2000.0	&	6.89		& G4V & \\
\end{supertabular}

\newpage
\begin{table}
 \centering
 \begin{minipage}{140mm}
 \caption{ 
Radial Velocities (RV) for HD~216437 are referenced to the Solar 
System barycentre but have an arbitrary zero-point determined by 
the radial velocity of the template. The JDs are topocentric.}
  \begin{tabular}{@{}lrr@{}}
JD&RV&Uncertainty \\
(-2450000)&(m~s$^{-1}$)&(m~s$^{-1}$) \\
& & \\
   830.9420  &   -38.8  &  4.6 \\
  1034.2251  &   -35.6  &  4.9 \\
  1386.3051  &    -3.4  &  5.7 \\
  1472.9552  &     4.5  &  4.1 \\
  1683.3146  &    36.2  &  4.8 \\
  1684.3276  &    34.5  &  4.2 \\
  1743.2343  &    50.9  &  6.1 \\
  1767.2046  &    37.9  &  4.3 \\
  1768.2248  &    33.5  &  5.3 \\
  1828.0427  &    42.8  &  5.2 \\
  1828.9634  &    40.9  &  5.1 \\
  1829.9568  &    43.7  &  5.7 \\
  1856.0478  &    30.6  &  7.4 \\
  1919.9294  &    28.9  &  5.3 \\
  1920.9255  &    29.2  &  6.6 \\
  2061.2882  &   -21.1  &  4.9 \\
  2092.2206  &   -15.1  &  5.2 \\
  2127.1981  &   -24.9  &  5.5 \\
  2154.1065  &   -29.9  &  4.5 \\
  2188.0807  &   -44.0  &  4.0 \\
  2387.3194  &   -22.0  &  4.2 \\
  2388.3068  &   -17.3  &  2.2 \\
  2389.2962  &   -22.7  &  6.6 \\
  2390.3183  &   -27.2  &  4.5 \\
  2422.3086  &   -27.9  &  4.0 \\
  2425.3260  &   -23.3  &  3.7 \\
\end{tabular}
\end{minipage}
\end{table}

\newpage
\begin{table}
 \centering
 \begin{minipage}{140mm}
 \caption{
Radial Velocities (RV) for HD~196050 are referenced to the Solar
System barycentre but have an arbitrary zero-point determined by
the radial velocity of the template. The JDs are topocentric.}
  \begin{tabular}{@{}lrr@{}}
JD&RV&Uncertainty \\
(-2451000)&(m~s$^{-1})$&(m~s$^{-1})$ \\
& & \\
   118.9450  &    10.9  & 10.7 \\
   411.0456  &    52.3  &  7.9 \\
   472.9298  &    57.6  &  5.5 \\
   683.1958  &    56.6  &  6.4 \\
   706.1291  &    51.7  &  7.3 \\
   743.0754  &    49.5  &  6.0 \\
   745.1895  &    46.7  &  5.3 \\
   767.0285  &    37.3  &  5.5 \\
   770.1480  &    43.3  &  6.6 \\
   827.9868  &    17.5  &  6.9 \\
   855.9770  &    24.0  &  9.3 \\
  1010.2975  &   -18.9  &  6.6 \\
  1061.1955  &   -21.9  &  5.7 \\
  1092.1221  &   -15.4  &  6.6 \\
  1127.1045  &   -27.7  &  6.0 \\
  1128.0595  &   -30.4  &  7.9 \\
  1130.0415  &   -22.1  &  5.7 \\
  1151.9802  &   -25.5  &  5.9 \\
  1153.8857  &   -36.6  &  4.7 \\
  1186.9195  &   -39.6  &  4.1 \\
  1187.9809  &   -45.2  &  3.5 \\
  1188.9390  &   -41.8  & 11.4 \\
  1189.9371  &   -32.9  &  5.3 \\
  1360.2972  &   -11.2  &  6.0 \\
  1387.3049  &    -2.1  &  4.7 \\
  1388.2519  &    -6.8  &  4.8 \\
  1389.2115  &    -6.3  &  5.3 \\
  1390.2928  &    -7.4  &  5.3 \\
  1421.2467  &    23.5  &  5.6 \\
  1425.2942  &    19.4  &  4.8 \\
\end{tabular}
\end{minipage}
\end{table}

\newpage
\begin{table}
 \centering
 \begin{minipage}{140mm}
 \caption{
Radial Velocities (RV) for HD~160691 are referenced to the Solar
System barycentre but have an arbitrary zero-point determined by
the radial velocity of the template. The JDs are topocentric.}
  \begin{tabular}{@{}lrr@{}}
JD&RV&Uncertainty \\
(-2450000)&(m~s$^{-1})$&(m~s$^{-1}$) \\
& & \\
   915.2911  &   -11.4  &  6.3 \\
  1118.8874  &   -10.2  &  3.4 \\
  1119.9022  &   -10.9  &  2.8 \\
  1120.8870  &   -11.2  &  3.0 \\
  1121.8928  &   -10.3  &  2.9 \\
  1236.2864  &   -26.5  &  4.0 \\
  1410.8977  &   -47.8  &  2.9 \\
  1412.9780  &   -47.7  &  5.6 \\
  1413.8981  &   -41.0  &  2.7 \\
  1630.3042  &    39.8  &  3.0 \\
  1683.0926  &    43.9  &  3.7 \\
  1684.1320  &    41.0  &  4.1 \\
  1718.1184  &    28.5  &  3.6 \\
  1742.9096  &    15.7  &  3.1 \\
  1743.9240  &    25.7  &  3.8 \\
  1745.0440  &    13.8  &  3.3 \\
  1766.9330  &     5.8  &  3.2 \\
  1767.9689  &     7.5  &  3.3 \\
  1827.8973  &    -5.9  &  2.9 \\
  1828.8866  &    -0.4  &  3.6 \\
  1829.8890  &    -6.8  &  3.4 \\
  1855.9058  &    -7.0  &  4.9 \\
  1984.2618  &   -20.5  &  3.5 \\
  2010.2829  &   -15.6  &  6.1 \\
  2061.1132  &   -27.7  &  3.1 \\
  2091.9807  &   -30.3  &  3.6 \\
  2126.9766  &   -12.0  &  3.7 \\
  2151.9693  &    -7.0  &  3.3 \\
  2152.9493  &     4.3  &  2.4 \\
  2153.8626  &    -0.9  &  2.8 \\
  2186.9095  &    20.9  &  2.6 \\
  2187.8879  &    16.7  &  2.7 \\
  2360.3245  &    34.6  &  2.3 \\
  2387.1722  &    34.6  &  2.7 \\
  2388.2097  &    31.0  &  2.9 \\
  2421.1696  &    15.8  &  2.7 \\
  2425.1226  &    20.3  &  3.1 \\
\end{tabular}
\end{minipage}
\end{table}

\begin{table}
 \centering
 \begin{minipage}{140mm}
  \caption{Orbital parameters for the companions to HD~216437, HD~196050
and HD~160691. The solution for HD~160691b is for the case of a single
Keplerian fit to the data whereas the fit for HD~160691c is based on
a two Keplerian fit. HD~160691c is uncertain so its best fit values
are shown in parentheses.}
  \begin{tabular}{@{}lrrrr@{}}
& HD~216437b & HD~196050b & HD~160691b & HD~160691c \\ 
Orbital Period (d) & 1294$\pm$250& 1288$\pm$230 &638$\pm$10&  (1300)\\
eccentricity & 0.33$\pm$0.09& 0.28$\pm$0.15& 0.31$\pm$0.08 & (0.8) \\
$\omega$ (degrees) & 79$\pm$30& 223$\pm$30& 320$\pm$30& (99) \\
Radial velocity semi-amplitude $K$ (m~s$^{-1}$) & 38$\pm$4& 41$\pm$5 &40$\pm$5 &(34.2) \\
Periastron Time (HJD)  &50681$\pm$200 & 51033$\pm$180&50958$\pm$30 & (51613)\\
M~sin~$i$ (M$_{\rm JUP}$) & 2.1$\pm$0.3 & 3.0$\pm$0.5& 1.7$\pm$0.2& (1)\\
a (au) & 2.4$\pm$0.5 & 2.3$\pm$0.5 & 1.5$\pm$0.1 & (2.3) \\
RMS residuals to fit (m~s$^{-1}$) & 6.64 & 7.51& 5.42 & (5) \\
\end{tabular}
\end{minipage}
\end{table}

\begin{table}
 \centering
 \begin{minipage}{140mm}
  \caption{Extra-solar planetary systems classified by orbital parameters of 
period and eccentricity (from http://exoplanets.org). 
The boundaries for classification are chosen
in terms of the Solar System, so eccentricity is chosen 
as 0.25 (cf. Pluto) and period as 88~d (cf. Mercury). Where
there is more than one planet present around a star the classification 
is made in terms of the inner planet. Outer planets in the system 
are recorded in the appropriate section though their entry is in italics 
to indicate that they are not included in the count.}
  \begin{tabular}{p{3.1cm}p{1cm}p{9.5cm}}
Class & Number & Objects \\
& & \\
${\rm 51~Peg~b -like}$ \hspace{0.8cm}  'circular short-period' \hspace{0.05cm}$e < 0.25, T < 88$ d & 24& 
HD~83443b, HD~46375b, HD~179949b, HD~187123b, Tau~Boo~b, 
BD~-103166b, HD~75289b, HD~209458b, 51~Peg~b, Ups~And~b, HD~49674b,
HD~68988b, HD~168746b, 
HD~217107b, HD~130322b, HD~38529b, 55~Cnc~b, GJ~86b, HD~195019b, 
Rho~Cr~Bb, GJ~876b, HD~121504b, HD~178911Bb, HD~16141b \\ 
& & \\
${\rm HD~114762-like}$ \hspace{0.6cm}  'eccentric short-period' $e > 0.25, T < 88$ d & 6& 
HD~162020b, HD~108147b, HD~6434b, {\em GJ~876c}, HD~74156b, HD~168443b, 
HD~114762b\\
& & \\
${\rm 70~vir~b -like}$ \hspace{1.1cm}'eccentric long-period' \hspace{0.05cm}$e > 0.25, T > 88$ d& 
37 & HD~80606b, 70~Vir~b, HD~52265b, HD~1237b, 
HD~37124b, HD~73526b, {\em HD~82943c}, HD~8574b, HD~169830b, 
HD~12661b, HD~89744b, HD~40979b,
HD~202206b, 
HD~134987b, 
HD~92788b, HD~142b, HD~177830b, HD~4203b, HD~210277b, HD~82943b, HIP~75458b, 
HD~222582b, HD~141937b, HD~160691b, HD~213240b, 16~Cyg~B~b, HD~196050b,
HD~114729b,
HD~190228b, HD~136118b, HD~50554b, HD216437b, {\em Ups~And~d}, {\em HD~12661c} 
HD~33636b, 
HD~106252b, HD~145675b, HD~72659b,
HD~39091b, {\em HD~38529c}, {\em HD~74156c}, Eps~Eri~b \\
& & \\
${\rm Solar~System -like}$ \hspace{0.6cm}    $e < 0.25, T > 88$ d& 13 & 
HD~37124b, {\em Ups~And~c}, HD~17051b, 
HD~28185b, HD~108874b, HD~128311b, HD~27442b, HD~19994b, 
HD~114783b, HD~23079b, HD~4208b, HD~10697b, 
47~Uma~b, HD~30177b, {\em 47~Uma~c}, {\em HD~168443c}, {\em 55~Cnc~c} \\
\end{tabular}
\end{minipage}
\end{table}

\label{lastpage}
\end{document}